# Superconducting plasmonics and extraordinary transmission

**A. Tsiatmas[1], R. Buckingham[2], V. A. Fedotov[1], S. Wang[3], Y. Chen[4], P. de Groot[2] and N. I. Zheludev[1]**

[1]Optoelectronics Research Centre, University of Southampton,
Southampton, Hampshire SO17 1BJ, United Kingdom
email: vaf@orc.soton.ac.uk

[2]School of Physics and Astronomy, University of Southampton,
Southampton, Hampshire SO17 1BJ, United Kingdom

[3]Shanghai Institute of Technical Physics, Chinese Academy of Sciences,
Shanghai, 200083, China

[4]Micro and Nanotechnology Centre, Rutherford Appleton Laboratory,
Didcot, Oxon OX11 0QX, United Kingdom

**Abstract**

Negative dielectric constant and dominant kinetic resistance make superconductors an intriguing plasmonic media. Here we report on the first study of one of the most important and disputed manifestations of plasmonics, the effect of extraordinary transmission through an array of sub-wavelength holes, using a perforated film of high-temperature superconductor.

The effect of extraordinary transmission can be regarded as one of the most important and disputable phenomenon in the area of plasmonics. It was observed as sharp peaks in transmission spectra of non-diffracting periodic arrays of sub-wavelength holes made in thin metal films [1]. The transmission efficiency at those maxima exceeded unity (when normalized to the area of the holes), which was orders of magnitude greater than predicted by standard aperture theory. Such unusual optical properties were attributed to resonant coupling of light with plasmons mediated by periodic patterning of the metal films.

Superconductors can be regarded as media with negative real dielectric constant and mainly kinetic resistance, and therefore electromagnetics of such structures falls into the domain of plasmonics. Here we report the first experimental data on observation of extraordinary transmission in a perforated non-diffracting superconducting film and demonstrate that the magnitude of the extraordinary transmission resonances can be controlled with temperature and increases dramatically upon superconducting phase transition.

Our experiments were performed using a free-space setup, which is based on mm-wave test system equipped with horn antennas and a closed-cycle liquid-helium cryostat, as shown in Fig. 1a and 1b. The superconduting film had the thickness of 300 nm and was made of high-temperature superconductor YBCO deposited on a low-loss sapphire substrate. It was perforated by etching an array of 1 mm holes with the period of 2.7 mm (see Fig. 1c), which rendered the structure non-diffracting at frequencies below 110 GHz. Transmission of the perforated cuprate film was measured for normal incidence at temperatures above and below its critical temperature $T_c$ = 87K in 75 – 110 GHz range of frequencies.

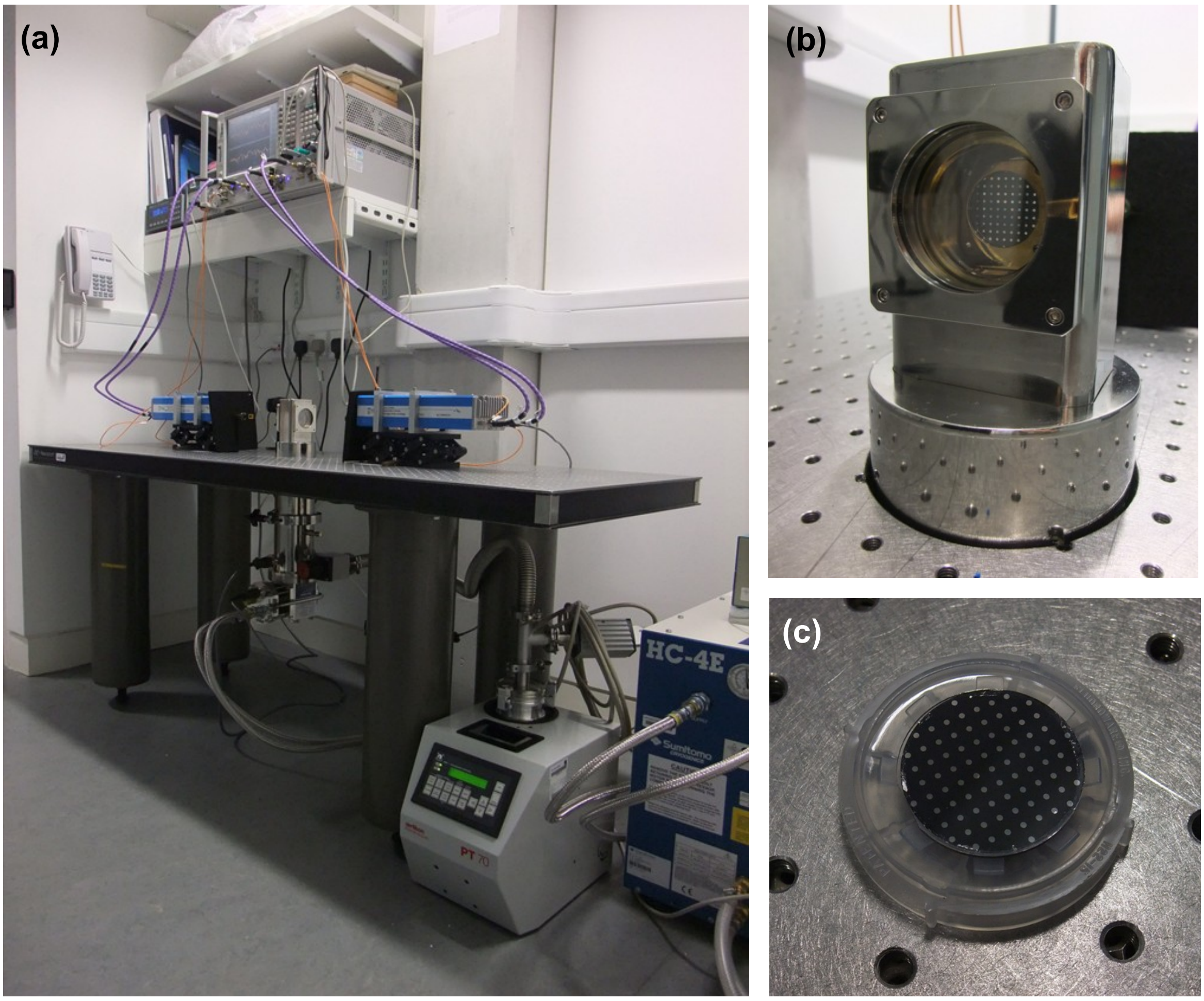

Fig. 1. (a) - Experimental mm-wave setup for investigating electromagnetic properties of structured superconducting films in free space. (b) - Head of a closed-cycle liquid-helium optical cryostat with mounted hole-array sample. (c) - Superconducting hole-array sample is made of YBCO film periodically perforated with sub-wavelength holes.

Our measurements clearly showed that the extraordinary transmission resonances exist even above the critical temperature of the structure and become substantially stronger upon superconducting phase transition. The results of the measurements are summarized on Fig. 2 in terms of changes in the transmission of the array relative to its room temperature state. In particular, Fig. 2a shows spectral response of the array at 100K, 70K and 2.6K, between 75 and 85 GHz where extraordinary transmission resonances can be seen as series of narrow peaks. It is evident that the spectral position and, more notably, the amplitude of the peaks change with temperature. On Fig. 2b we present the temperature dependence of the transmission change for a resonant peak located near 81 GHz. It shows that the amplitude of the resonance slowly increases as the temperature of the sample departs from 300K towards $T_c$, where a dramatic and abrupt increase of the sample transmission can be seen. Surprisingly, for temperatures below 60K the transmission change appears to decrease with temperature.

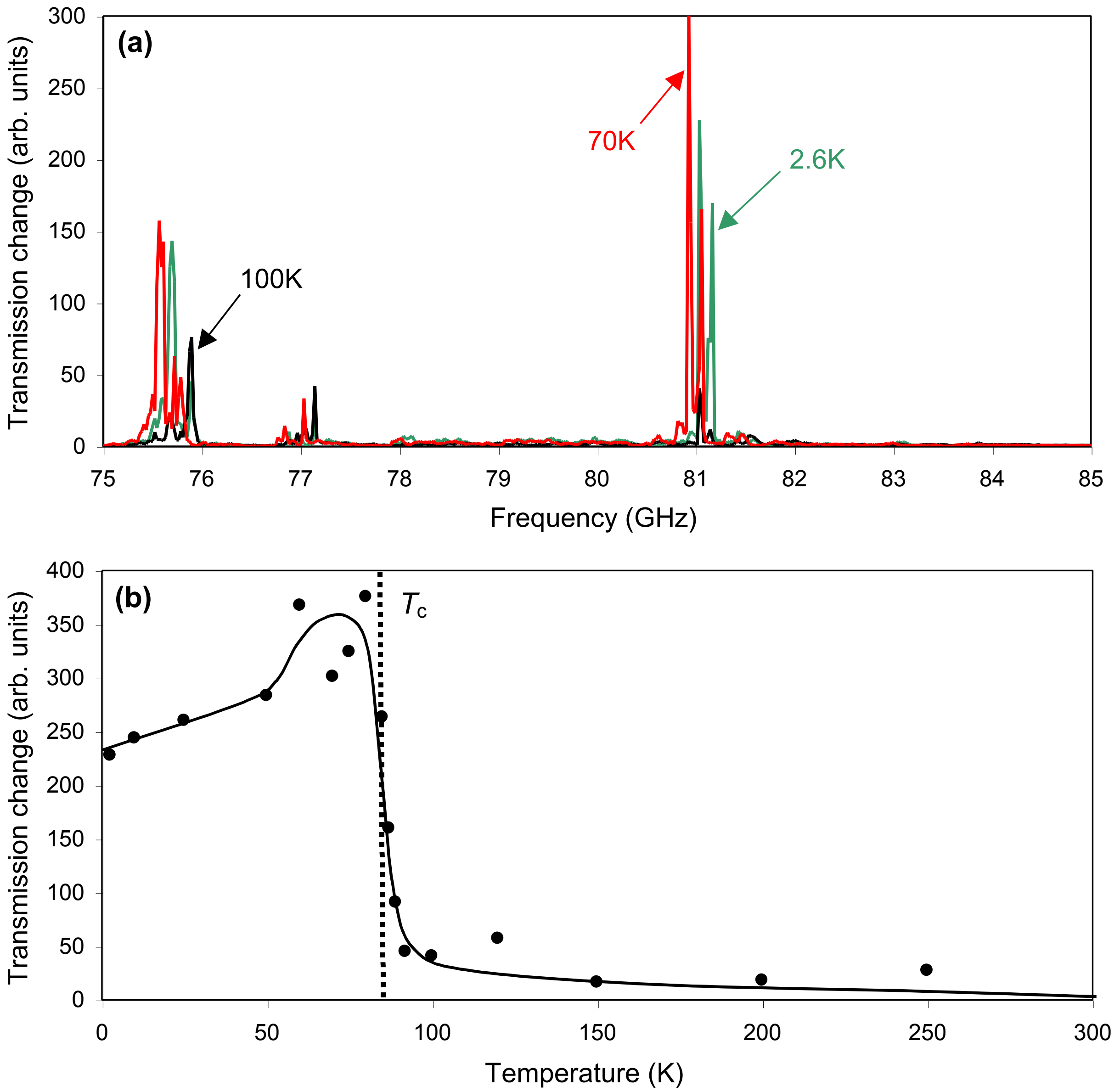

Fig. 2. Extraordinary transmission through YBCO hole-array. (a) - Spectra of the sample's transmission change relative to its room temperature state at three different temperatures. (b) - Change in the amplitude of the extraordinary transmission peak near 81 GHz as a function of the sample's temperature.

## References

[1] T. W. Ebbesen, H. J. Lezec, H. F. Ghaemi, T. Thio, P.A.Wolff, Extraordinary optical transmission through sub-wavelength hole arrays, *Nature*, 391, pp. 667-669, 1998